\documentclass[11pt,a4paper]{article}
\usepackage{jheppub}
\usepackage{lingmacros}
\usepackage{tree-dvips}
\usepackage{graphicx}
\usepackage{amsmath}
\usepackage{amssymb}
\usepackage{latexsym}
\usepackage{color}
\usepackage{commath}
\usepackage{braket}
\usepackage{mathtools}
\usepackage{dcolumn}
\usepackage[latin1]{inputenc}
\usepackage{graphicx, psfrag}
\usepackage{float}
\usepackage{tensor}
\usepackage{amsfonts}
\usepackage{hyperref}
\usepackage{subfigure}
\usepackage{pgfplots}
\usepackage{epstopdf}
\usepackage{booktabs}
\usepackage{enumitem}

\title{Solid/Liquid Phase Transition and Heat Engine in Asymptotically Flat Schwarzschild Black Hole via the R\'enyi Extended Phase Space Approach}

\author[a,b]{Chatchai Promsiri,} 
\author[b,c]{Ekapong Hirunsirisawat}
\author[a,b,d]{and Watchara Liewrian}

\affiliation[a]{Department of Physics, Faculty of Science, King Mongkut's University of Technology Thonburi, Pracha Uthit Road, Bangkok, 10140, Thailand}
\affiliation[b]{Theoretical and Computational Physics (TCP), KMUTT, Theoretical and Computational Science Center(TaCS), Faculty of Science, King Mongkut's University of Technology Thonburi, Pracha Uthit Road, Bangkok, 10140, Thailand}
\affiliation[c]{Learning Institute, King Mongkut's University of Technology Thonburi, Pracha Uthit Road, Bangkok, 10140, Thailand}
\affiliation[d]{Thailand Center of Excellence in Physics, Ministry of Higher Education, Science, Research and Innovation, 328 Si Ayutthaya Road, Bangkok 10400, Thailand}

\emailAdd{chatchaipromsiri@gmail.com} 
\emailAdd{ ekapong.hir@mail.kmutt.ac.th}
\emailAdd{watchara.liewrian@mail.kmutt.ac.th}

\abstract{Recently, it has been found that, with the R\'enyi statistics, the asymptotically flat Schwarzschild black hole can be in thermal equilibrium with infinite heat reservior at a fixed temperature when its event horizon radius is larger than the characteristic length scale $L_\lambda=1/\sqrt{\pi \lambda}$, where $\lambda$ is the nonextensivity parameter.  In the R\'enyi extended phase space with the $PdV$ work term,  an off-shell free energy in the canonical ensemble with the thermodynamic volume as an order parameter is considered to identify a first-order Hawking-Page (HP) phase transition as a solid/liquid phase transition.  It has the latent heat of fusion from solid (corresponding to thermal radiation) to liquid (corresponding to black hole) in the form of  $\sim 1/\sqrt{\lambda}$; this is evident of the absence of the HP phase transition in the case of asymptotically flat Schwarzschild black hole from the GB statistics ($\lambda=0$).  Moreover, we investigate the generalized second law of black hole thermodynamics (GSL) in R\'enyi statistics by considering the black hole as a working substance in heat engine. Interestingly, an efficiency $\eta$ of the black hole in a Carnot cycle takes the form $\eta_c=1-T_\text{C}/T_\text{H}$.  This confirms the validity of the GSL in the R\'enyi extended phase space.}

\keywords{phase transition, black hole, heat engine, R\'enyi statistics}

\begin{document}
\maketitle

\section{Introduction}

The notion of a black hole having the nature of thermal object originated from the surprising mathematical parallels between the laws of black hole mechanics and of thermodynamics~\cite{Bardeen}.  Since then, the geometrical properties of black hole's event horizon has been found to be linked with its thermodynamic properties.  Bekenstein was the first to postulate that the surface gravity and the area of a black hole's event horizon could be regarded as the temperature and the entropy of black hole, respectively~\cite{Bekenstein1}.  While the geometrical considerations through general relativity cannot give a black hole any thermal concept, the Hawking's consideration on the quantum effect around the event horizon predicts the emission of thermal radiations at certain temperature to an observer at infinity~\cite{Hawking1}.  The need of taking into account the quantum properties into black holes may imply that the black hole thermodynamics may be related in some ways with the quantum theory of gravity.

Recently, there are some concerns whether the conventional thermodynamics, based on the Gibbs-Boltzmann (GB) statistics, is valid in deriving the thermodynamic properties of the extreme situation like black hole.  It is essential to elaborate here the important issues of black hole thermodynamics from the GB statistical approach, which could not be resolved satisfactorily. They are as follows:
\begin{enumerate}[label=(\roman*)]
	\item The GB statistics is appropriate to describe only a weakly interacting system whose the entropy is extensive.  Nevertheless, the black hole system has long-range gravitational interactions and, according to some clues, may have non-local quantum correlations among its constituents, such that it should not be an extensive system.  This  is evident through the area law of its entropy~\cite{Bombelli, Srednicki, Eisert}. Therefore, a black hole is nonextensive system which cannot be described well using the thermodynamics based on the GB statistics.
	\item Based on the GB statistics, a black hole surrounded by an infinite bath of thermal radiation in an asymptotically flat spacetime is unstable due to the presence of its negative heat capacity. This implies that the canonical ensemble in the GB statistics cannot be applied in the case of nonextensive long-range interaction black hole system in asymptotically flat spacetime at a fixed temperature \cite{Hawking2}, for a good review, see \cite{Padmanabhan}.
	\item Basically, the volume within black hole's event horizon can expand when absorbs some energy, whereas it can shrink as it emits the Hawking's radiation.  One may wonder whether the thermodynamic system corresponding to the black hole can perform the mechanical work in a similar way as occuring in a gas system in the GB thermodynamics.  Unfortunately, the notion of  thermodynamic pressure and volume are absent in the standard black holes physics. Therefore, the first law of black hole thermodynamics does not include the mechanical work $PdV$ term to describe the change of internal energy in thermodynamic processes.    
\end{enumerate}

To address these isssues, we may need to use an alternative choice of the generalized entropy function.  Recently, the Tsallis~\cite{Tsallis1} and R\'enyi~\cite{Renyi} entropies, which is based on the power law probability distribution, has been used to study strongly interacting phenomena in particle physics~\cite{BiroBarVan, Wong, Deppman}, kinetic theory~\cite{Lima, Biro1, Mitra} and complex systems~\cite{Thurner}. A comprehensive description of generalized entropies and its applications has been found in \cite{Tsallis2}.  In black hole physics, its thermodynamic nature was firstly explored via the nonextensive extropy by Tsallis~\cite{Tsallis3}. The thermodynamic stability and phase structure of the Schwarzschild, Kerr and Reissner-Nordstr\"om black holes in an asymptotically flat spacetime were investigated via R\'enyi statistics in~\cite{Czinner1, Czinner2, Chatchai2}. These two nonextensive entropies have the parameter $\lambda$, which is so called the nonextensive parameter.  The deviation of its value from zero quantifies how much the system's behaviors deviate from the extensive GB statistics, corresponding to $\lambda=0$.   

Studying with the R\'enyi staticstics, the black holes in asymptotically flat spacetime can be in stable thermodynamic equilibrium with thermal bath at a fixed temperature as $\lambda$ above zero.  Hence, the canonical ensemble of this system exists in a case of nonextensive R\'enyi statistics.  The role of nonextensive parameter $\lambda$ to stabilize the black holes is in a similar fashion as the gravitational potential with negative cosmological constant $\Lambda$ in AdS space. Moreover, the nonextensive effect can stabilize the Schwarzschild black hole in a de Sitter spacetime of the positive cosmological constant as well~\cite{Tannukij, Rath}.

As mentioned above in (iii), the notion of thermodynamic pressure and volume of black holes are absent such that some thermodynamic behaviors of black hole are not be allowed to mathenatically match with those in the thermodynamics of conventional matter.  For instance, even though it has been found that the Van der Waals (VdW) like phase transition exists in the $q-\phi$ plane of the AdS Reissner-Nordstr\"om black hole (RN-AdS) in a canonical ensemble, we cannot exhibit this phase transition in the $P-V$ plane~\cite{Chamblin1, Chamblin2, Niu, Chatchai}.   A more complete analogy between the conventional matter and a black hole has been established by introducing a negative cosmological constant $\Lambda$ as a thermodynamic variable, namely the thermodynamic pressure $P=-\frac{\Lambda}{8\pi}$.  Its conjugate is the thermodynamic volume $V=\frac{4}{3}\pi r_\text{h}^3$, where $r_\text{h}$ is an event horizon radius.  In this approach, the mass turns out to be an enthalpy which can be written as $M=H=U+PV$~\cite{Kastor, Mann1}. This framework is known as black hole thermodynamics in the extended phase space or black hole chemistry (for review see \cite{Kubiznak, Mann2, Mann3}, and references therein). 

Without the negative cosmological constant, the asyptotically flat black hole for example has the vanishing pressure, and hence the notion of thermodynamic pressure and volume are absent.  Fortunately, the authors in \cite{Chatchai2} have proposed an idea that the nonextensive parameter $\lambda$ can be regared as a thermodynamic variable.   As a result, $\lambda$ is allowed to play a role of the thermodynamic pressure. In this way, we identified the thermodynamic pressure $P=\frac{3\lambda}{32}$ and its conjugate variable as the thermodynamic volume $V=\frac{4}{3}\pi r_\text{h}^3$. In addition, the consistent Smarr formula from the scaling argument have revealed that the black hole mass $M$ should be interpreted as an enthalpy instead of the internal energy of the black hole.  This is in the same way as the extended phase space in AdS, as mentioned above. This  framework is called the R\'enyi extended phase space approach, or may be dubbed the R\'enyi black hole chemistry. 

Intriguingly, it has been found in \cite{Chatchai2} that there exists a VdW like phase transition of the RN-flat in this framework.  One may wonder whether there exists some interesting phase transitions in the asymptotically flat Schwarzschild black hole (Sch-flat) from the R\'enyi extended phase space approach.  From a point of view, applying this in a variety of cases is necessary for testing its validity and exploring some interesting aspects of this new approach.   To acheive this, we consider in this paper the Gibbs free energy, using this appraoch, to investigate a first order radiation/large black hole phase transition in the asymptotically flat background.   

Moreover, it is interesting to investigate whether the thermodynamic process from this setup obey the generalized second law of thermodynamics (GSL), which states that the sum of black hole entropy and that of the environment outside the horizon never decreases~\cite{Bekenstein2}. The GSL is stronger than a classical area theorem \cite{Hawking3} because it is also valid when the Hawking radiation is taken into account. A black hole emits particles via the process of Hawking radiation. Consequently, the area of its event horizon becomes smaller implying a decrease in black hole's entropy. This semi-classical process has therefore violated the classical area theorem and also the second law of thermodynamics. However, the GSL is still valid despite the decrease of horizon area;  the increase in the entropy of the environment occupied by Hawking radiations emitted from the black hole overcomes the decrease of Bekenstein-Hawking entropy such that the overall entropy keeps growing up.

There are several attempts to test the validity of the GSL, for a review see \cite{Wald, Wall}. In the context of the extended phase space approach, the holographic heat engine in the $P-V$ plane through considering the AdS black hole as a working substance was proposed by Johnson \cite{JohnsonI, ChakrabortyI, ChakrabortyII}. Several holographic heat engines have been studied later in \cite{JohnsonII, JohnsonIII, Chandrasekhar, Mo, Zhang}. Inspired by Johnson's work, the thermal efficiency of black hole heat engine will be derived in the present paper to investigate the GSL validity from the macroscopic point of view instead of the notion of information theoretic and statistical arguments. From the Carnot statement that ``no heat engine, operating between the same high and low temperatures, can be more efficient than a Carnot engine''~\cite{Reichl},  the upper bound of the efficiency coefficient is in the form $\eta_c=1-\frac{T_\text{C}}{T_\text{H}}$. In other words, the existence of heat engine with $\eta >\eta_c$ violates the second law of thermodynamics.  Applying the R\'enyi extended phase space in the Sch-flat can give the corresponding reversible thermodynamic processes in a heat engine, which can be defined as a closed path in the $P-V$ diagram. The presence of the $PdV$ term allows us to consider the mechnical work from the expansion and contraction of black hole, in the same way as conventional matter.  Thus, apart from exploring the phase transition, we consider in the present work a Carnot cycle in the $P-V$ diagram and determine a thermal efficiency coefficient of the black hole heat engine.  This is an additional validity test of the R\'enyi extended phase space approach in black hole thermodynamics to ensure that the GSL is satisfied.

This paper is organized as follows.   We begin with the review of nonextensive statistics, namely the Tsallis and R\'enyi statistcis, and the deep reasons why we need it in the consideration of black hole thermodynamics in section \ref{Tsallis-R stat}. Next, the GB and R\'enyi thermodynamics of Sch-flat are reviewed in section \ref{R-stat&BH thermo}.  The thermodynamic behaviors of black hole from the R\'enyi entropy are also discussed in this section.  Moreover, the formulation of black hole thermodynamics of Sch-flat in the R\'enyi extended phase space is derived using the generalized Smarr formula in section \ref{Generalized Smarr Formula}.   With this approach, the Hawking-Page phase transition of the Sch-flat black hole corresponding to the solid/liquid phase transition in conventional matter and its latent heat are discussed in section \ref{Phase Transition}.  Importantly, the cyclic thermodynamic process of the Sch-flat from the R\'enyi statistics, namely the balck hole heat engine, is discussed in section \ref{BH heat engine}.  We end up with the conclusion in section \ref{Conclusion}.

\section{Tsallis and R\'enyi Statistics} \label{Tsallis-R stat}
 
A non-extensive generalization of GB entropy was proposed by Tsallis~\cite{Tsallis1}. For a system of $W$ discrete states, the Tsallis entropy can be written in the form
\begin{eqnarray}
S_\text{T} = \frac{1}{1-q}\left( \sum _{i=1}^W p_i^q-1 \right), \label{Tsallis}
\end{eqnarray}
where $p_i$ is the probability associated to a state $i$ and $q\in \mathbb{R}$ is a real parameter.   Defining a non-extensivity parameter $\lambda \equiv 1-q$, the above formula becomes the GB entropy when $\lambda \rightarrow 0$.  It is noteworthy here  to indicate that the GB entropy is associated with the extensive system.  Considering a combined system AB consisting of the subsystems A and B, its entropy is extensive when it can be written in the form $S(W_\text{AB})=S(W_\text{A})+S(W_\text{B})$, where $W_\text{A}$, $W_\text{B}$, and $W_\text{AB}$ are the number of states of the systems A, B and AB, respectively.  Note that $W_\text{AB}=W_\text{A} W_\text{B}$ when A and B are independent.  However, the formula of entropy in \eqref{Tsallis} is in the non-extensive form, which is relaxed to a weaker non-addditive composition rule \cite{Abe}.  The Tsallis entropy of a combined system can be written as
\begin{eqnarray}
S_\text{T}(W_\text{AB}) = S_\text{T}(W_\text{A})+S_\text{T}(W_\text{B})+\lambda S_\text{T}(W_\text{A})S_\text{T}(W_\text{B}). \label{nonAd}
\end{eqnarray}
Note that this composition rule becomes additive when $\lambda=0$, corresponding to that of the GB entropy. However, the composition rule of the Tsallis entropy turns out to be in the additive form using an appropriate transformation.  We will be back to this point later.

\begin{figure}
	\centering
	\includegraphics[scale=0.48]{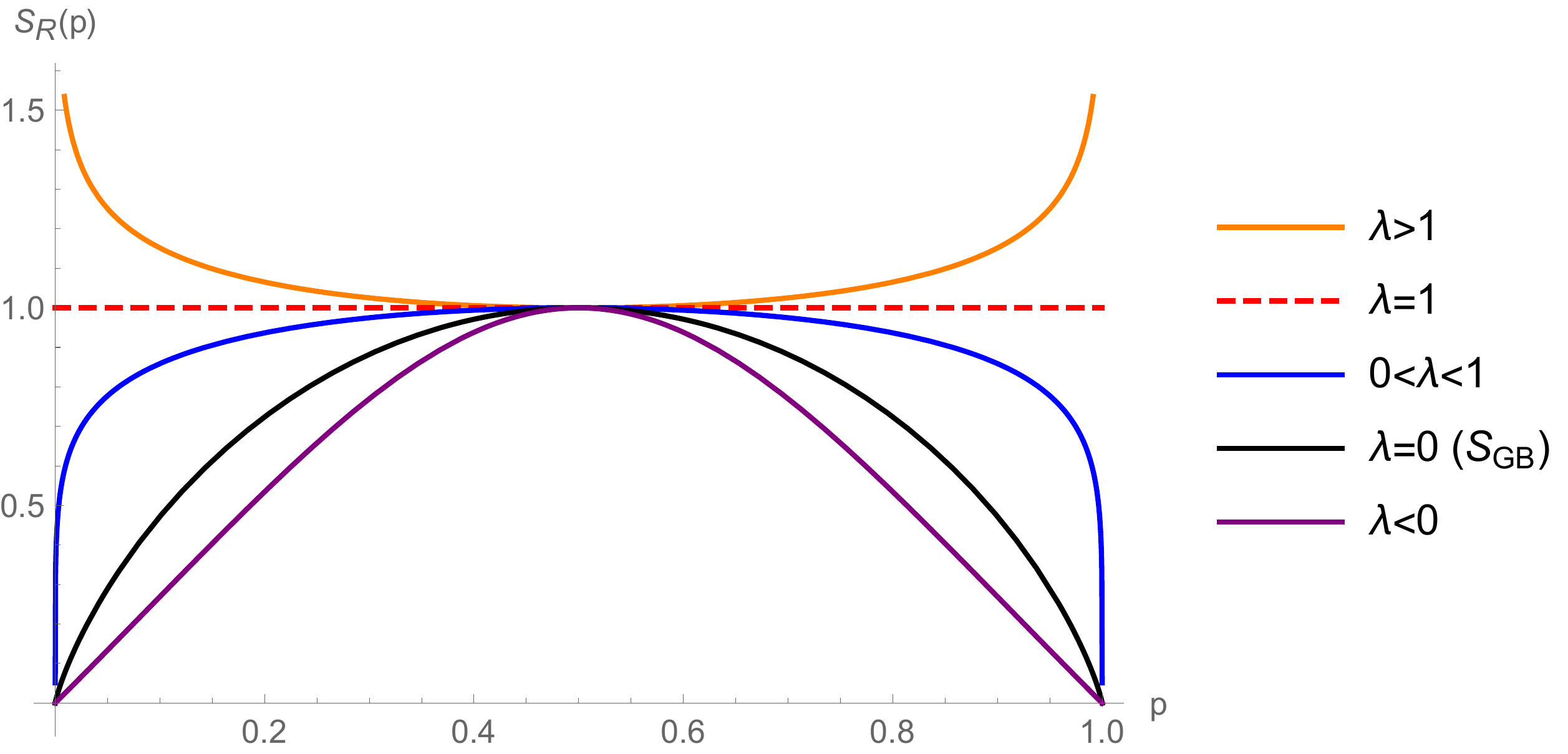}
	\caption{The R\'enyi entropy $S_\text{R}$ of the system of a biased coin with a random variable of two possible outcomes, namely heads or tails, is plotted as a function of the probability of landing on heads $p$ at different values of $\lambda$. Note that, at all values of $\lambda\leq 1$, the R\'enyi entropy  becomes to 1 when $p=0.5$, {\it i.e.}  a fair coin due to its maximum of surprisal.}\label{fig:1}
\end{figure}  

Here, let us introduce another generalized entropy with $\lambda$ parameter, the R\'enyi entropy~\cite{Renyi}, which is defined as
\begin{eqnarray}
S_\text{R} = \frac{1}{\lambda}\ln \sum _{i=1}^W p_i^{1-\lambda}. \label{Renyi}
\end{eqnarray}
Remark that the non-extensivity parameter $\lambda$ of the R\'enyi entropy has to be not more than 1, otherwise the entropy function becomes not well-defined due to its convexity.  This issue can be illustrated more through considering a biased coin with the probability of landing on heads and tails is $p$ and $1-p$, respectively.  Using \eqref{Renyi}, the R\'enyi entropy of this system can be written in the form 
\begin{eqnarray}
S_\text{R}(p) = \frac{1}{\lambda}\ln \left[ p^{1-\lambda}+(1-p)^{1-\lambda} \right].
\end{eqnarray}
As shown in \figref{fig:1},  the entropy function of $\lambda>1$ (orange,solid) is very large when $p\to 0$ or $1$.  This is not well-behaved since at $p\to 0$ ($p\to 1$), corresponding to that the coin  is highly probable to land on tails (heads), the outcome should have very small value of entropy.  Namely, the entropy of a biased coin should be maximum at $p=0.5$ (fair coin) and vanishes at $p=0$ and $1$, as can be seen in the plots of the entropy function with $\lambda=1$ (red, dashed), $0<\lambda <1$ (blue, solid), $\lambda=0$ (black, solid), and $\lambda < 0$ (purple, solid).  In other words, a well-defined entropy function needs to satisfy the concavity property of entropy~[ref].  In the present work on black hole thermodynamics with R\'enyi statistics, the parameter $\lambda$ in range  $0<\lambda<1$  is of our interest due to its good thermodynamic behavior as shown in \cite{Chatchai2}.

Considering  \eqref{Tsallis} and \eqref{Renyi}, it can be shown that the R\'enyi entropy $S_\text{R}$ can be written in term of the Tsallis entropy $S_\text{T}$ as
\begin{eqnarray}
S_\text{R} = \frac{1}{\lambda}\ln \left( 1+\lambda S_\text{T} \right). \label{RenyiTsallis}
\end{eqnarray}
From this transformation, the R\'enyi entropy of the combined system AB should be written in the form
\begin{eqnarray}
S_\text{R}(W_{\text{AB}}) = \frac{1}{\lambda}\ln \left( 1+ \lambda S_\text{T}(W_{\text{AB}}) \right).
\end{eqnarray}
By substituting the non-additive composition rule \eqref{nonAd} for $S_\text{T}$, we have
\begin{eqnarray}
S_\text{R}(W_{\text{AB}}) &=& \frac{1}{\lambda}\ln \left[ 1+ \lambda \left(S_\text{T}(W_{\text{A}})+S_\text{T}(W_{\text{B}})+\lambda S_\text{T}(W_{\text{A}})S_\text{T}(W_{\text{B}}) \right) \right] \nonumber \\
&=& \frac{1}{\lambda} \ln \left[ \left( 1+\lambda S_\text{T}(W_{\text{A}}) \right)\left( 1+\lambda S_\text{T}(W_{\text{B}}) \right) \right] \nonumber \\
&=& \frac{1}{\lambda} \ln \left( 1+\lambda S_\text{T}(W_{\text{A}}) \right)+ \frac{1}{\lambda} \ln \left( 1+\lambda S_\text{T}(W_{\text{B}}) \right) \nonumber \\
&=& S_\text{R}(W_{\text{A}})+S_\text{R}(W_{\text{B}}).
\end{eqnarray}
Therefore, the transformed entropy function, \textit{i.e.} the R\'enyi entropy, becomes to have additivity, although it is obtained from the non-additive Tsallis entropy \cite{Biro2}.

In thermal equilibrium of an isolated system with two subsystems, the total entropy has the maximum value, satisfying the condition $dS = dS_\text{A}+dS_\text{B}=0$.  According to the zeroth law of thermodynamics, the empirical temperature of these two subsystems are equal, and thus it can be defined via the relation $ T^{-1}=\frac{\partial S_\text{A}}{\partial E_\text{A}}=\frac{\partial S_\text{B}}{\partial E_\text{B}}$.  This indicates that the lack of additive form in nonextensive entropy tends to obscure the notion of  thermal equilibrium and empirical temperature. Since  the R\'enyi entropy is additive, we therefore have the entropy compatible with the zeroth law of thermodynamics for describing a non-extensive system. Hence, the empirical temperature function can be obtained from the usual relation
\begin{eqnarray}
\frac{1}{T_\text{R}}=\frac{\partial S_\text{R}}{\partial E}. \label{TR_1st}
\end{eqnarray}

Recently, it has been argued that the Bekenstein-Hawking entropy $S_{\text{bh}}$ of a black hole can be associated with the non-extensive Tsallis entropy \cite{Tsallis3}. As mentioned above, concerning about the zeroth law compatibility, it is thus appropriate to work on black hole thermodynamics using the R\'enyi entropy as a function the Bekenstein-Hawking entropy in the form
\begin{eqnarray}
S_\text{R} = \frac{1}{\lambda}\ln (1+\lambda S_{\text{bh}}). \label{Srbh}
\end{eqnarray}
Using \eqref{TR_1st} and \eqref{Srbh}, we can write the R\'enyi temperature $T_R$ in term of the Hawking temperature $T_\text{h}$
\begin{eqnarray}
T_\text{R}=T_\text{h} (1+\lambda S_{\text{bh}}) . \label{T_R-T_Hawk}
\end{eqnarray}
The thermodynamic properties and phase transitions in  several kinds of  black hole solutions with the R\'enyi entropy have 
been studied recently in \cite{Czinner1, Czinner2, Tannukij, Chatchai2}. 

\section{R\'enyi Statistics and Thermodynamics of Black Holes} \label{R-stat&BH thermo}

In this section, we review the standard GB and R\'enyi thermodynamics of a Schwarzschild black hole in asymptotically flat spacetime. The line element of spherically symmetric black holes can be written in the form
\begin{eqnarray}
ds^2 = -f(r)dt^2 + \frac{dr^2}{f(r)} + r^2(d\theta^2+\sin^2\theta d\phi^2).
\end{eqnarray}
For the Schwarzschild black hole solution, we have
\begin{eqnarray}
f(r) = 1-\frac{2M}{r},
\end{eqnarray}
where the parameter $M$ describes the mass of black hole. At the black hole horizon, we have $f(r_\text{h})=0$, where $r_\text{h}$ is the location of its event horizon. Consequently, the thermodynamic quantities can be written in the black hole's event horizon $r_\text{h}$ as
\begin{eqnarray}
M &=& \frac{r_\text{h}}{2}, \\
T_\text{h} &=& \frac{f'(r_\text{h})}{4\pi} = \frac{1}{4\pi r_\text{h}}, \\
S_{\text{bh}} &=& \frac{\mathcal{A}}{4} = \pi r_\text{h}^2, \\ 
C &=& T_\text{h}\left( \frac{\partial S_{\text{bh}}}{\partial T_\text{h}} \right) = -2\pi r_\text{h}^2,
\end{eqnarray}
where $T_\text{h}$ is the Hawking temperature, $S_{\text{bh}}$ is the Bekenstein-Hawking entropy which depend on the surface area of event horizon $\mathcal{A}$. Note that, with the GB statistics, the heat capacity $C$ has negative value implying that the asymptotically flat Schwarzschild black hole is thermodynamically unstable.

From \eqref{Srbh} and \eqref{T_R-T_Hawk}, the R\'enyi entropy and its corresponding Hawking temperature of the asymptotically flat Schwarzschild black hole takes the form
\begin{eqnarray}
S_\text{R} = \frac{1}{\lambda}\ln (1+\lambda \pi r_\text{h}^2) = \pi L_\lambda^2\ln \left(1+\frac{r_\text{h}^2}{L_{\lambda}^2}\right), \label{SrS}
\end{eqnarray}
and

\begin{figure*}[!ht]
	\begin{tabular}{c c}
		\includegraphics[scale=0.31]{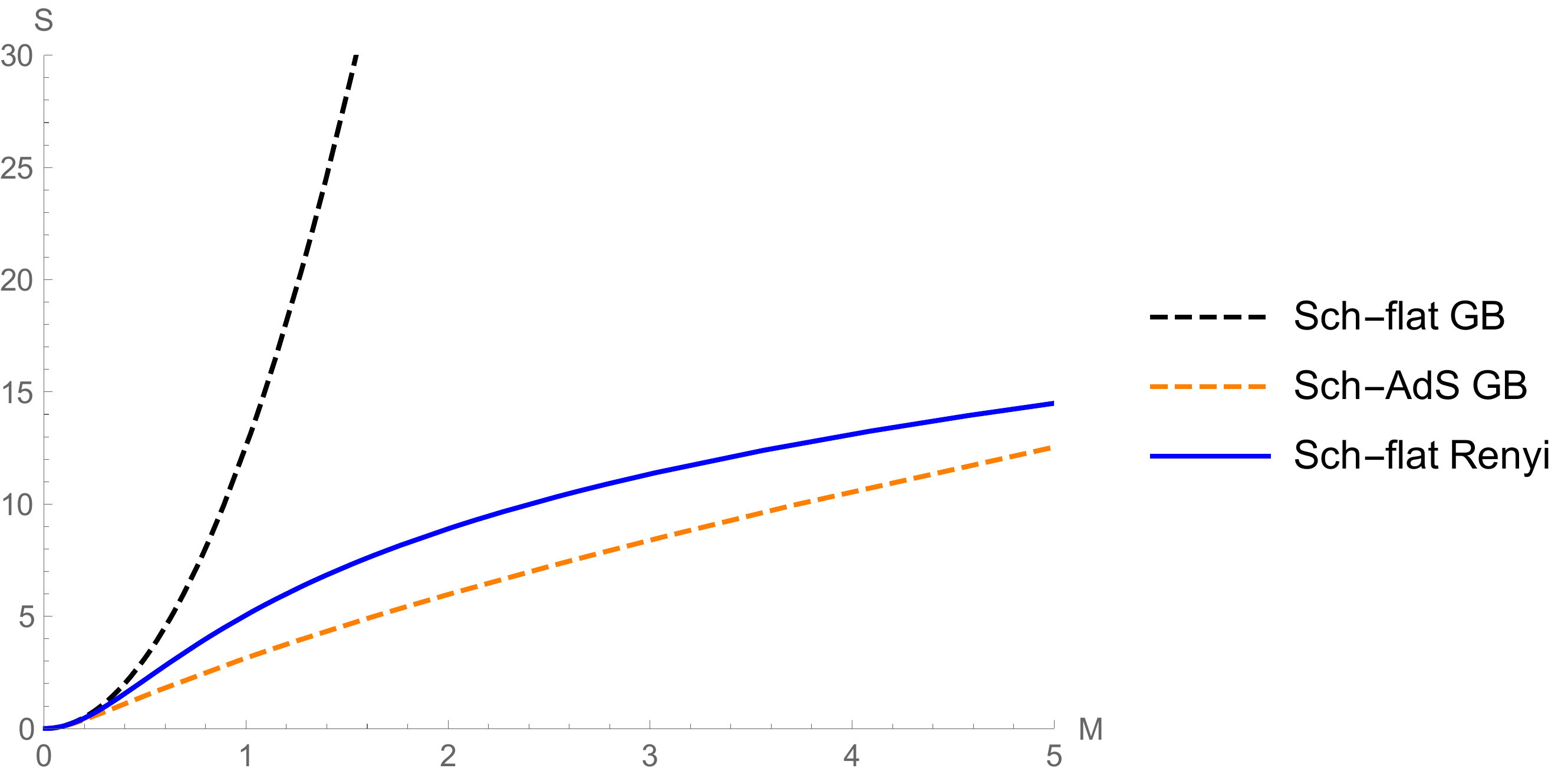}\quad
		\includegraphics[scale=0.31]{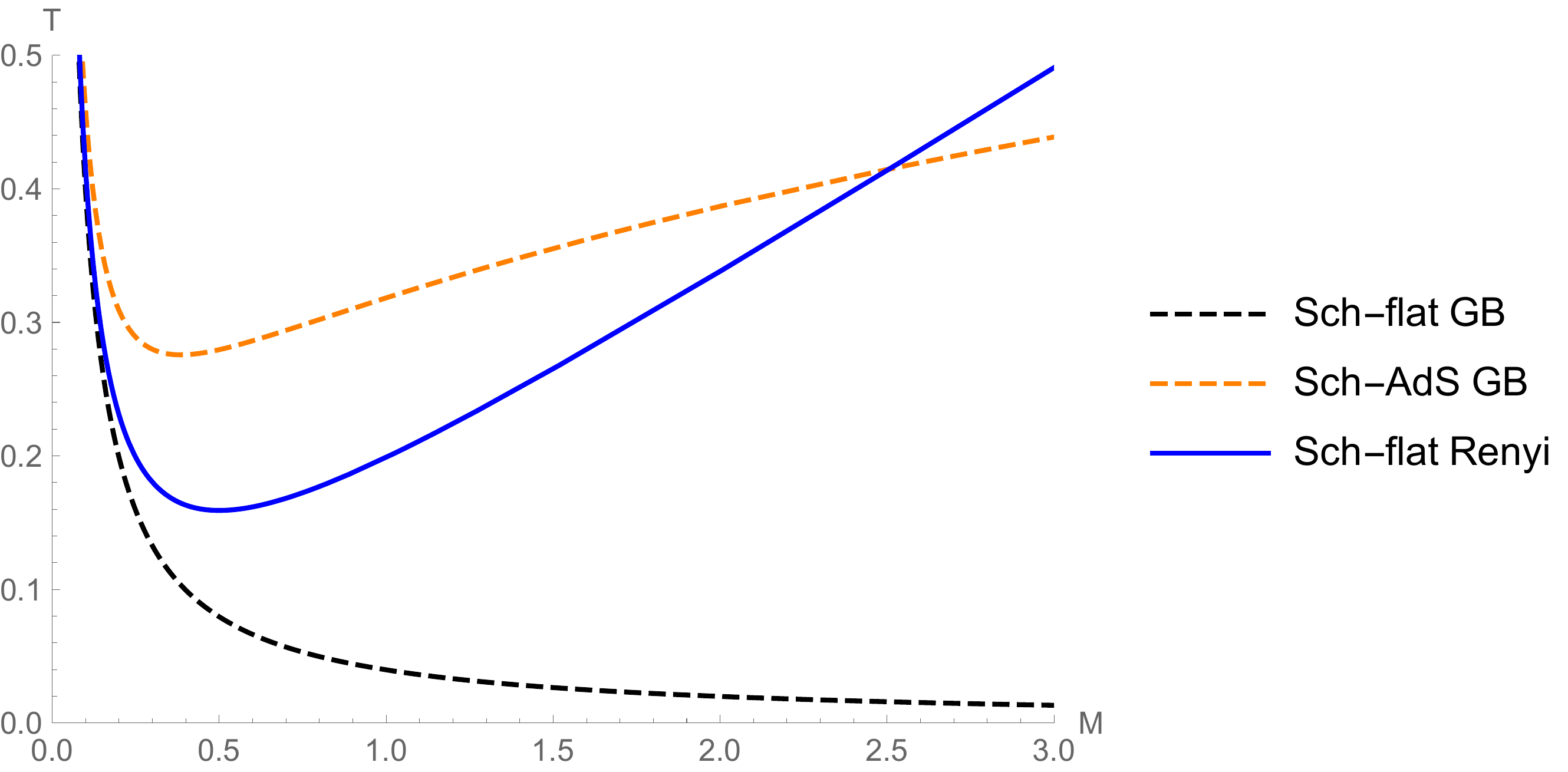}
	\end{tabular}
	\caption{Left: Entropy versus the mass (energy) of a black hole is plotted for the Sch-flat (dashed black) and the Sch-AdS (dashed orange) cases from the GB statistics, compared with one for the case of the Sch-flat from the R\'enyi statistics $L_\lambda =1$ (solid blue). Right: The figure shows the temperature versus the mass (energy) relation of the black hole. Unlike the Sch-flat in GB statistics case, the thermal behavior of black hole in R\'enyi framework has an interesting similarity to the AdS black hole. In the R\'enyi approach, this result reveals that the black hole bahaves like it is placed in the background with nonzero energy density, in the same way as ocurring in the AdS black hole system.}\label{fig:6}
\end{figure*}

\begin{eqnarray}
T_\text{R} = \frac{1}{4\pi r_\text{h}}(1+\lambda \pi r_\text{h}^2) = \frac{1}{4\pi r_\text{h}}\left( 1 + \frac{r_\text{h}^2}{L_{\lambda}^2} \right), \label{TrS}
\end{eqnarray}
where we have defined the characteristic length $L_\lambda=1/\sqrt{\pi \lambda}$.  Obviously, this nonextensivity length scale emerges as a result from the R\'enyi statistics; $L_\lambda$  becomes infinite when we back to the GB statistics, as $\lambda$ approaches zero.  For comparison, we plot the entropy and temperature versus the mass (energy) of black holes for the Sch-flat and the Sch-AdS in standard approach and those for the Sch-flat in the alternative R\'enyi statistics, as shown in \figref{fig:6}. Note that $\lambda$ is a dimensionless parameter, we perform the dimensional analysis of \eqref{SrS} to obtain a physical nonextensivity length scale. The physical length scale is given by $L_\lambda = l_P/\sqrt{\lambda \pi}$ where $l_P$ is the Planck length.  Remarkably, the nonextensivity length scale should be significantly larger than the Planck length as the quantum gravity effect is not included. Hence, we have 
\begin{eqnarray} \label{planck length}
	L_\lambda = \frac{l_P}{\sqrt{\lambda \pi}} > l_P, 
\end{eqnarray}
which provides the upper bound of the nonextensive parameter due to quantum gravity limit, {\it i.e.}
\begin{eqnarray} \label{lambda_bound}
	\lambda < \frac{1}{\pi}.
\end{eqnarray}
Thus, it makes sense to consider black hole thermodynamics from the R\'enyi statistics with the value of $\lambda$ large enough to see an effect of the deviation from the GB thermodynamics, but sufficiently small to be safe to discuss in semiclassical gravity framework without reaching quantum gravity limit.
Returning to the geometric units, the heat capacity of black hole can be obtained from
\begin{eqnarray}
C_\text{R} &=& T_\text{R}\left( \frac{\partial S_\text{R}}{\partial T_\text{R}} \right) = -\frac{2\pi r_\text{h}^2}{1-\lambda \pi r_\text{h}^2} = - \frac{2\pi r_\text{h}^2}{1-\frac{r_\text{h}^2}{L_{\lambda}^2}}. \label{c}
\end{eqnarray}
Notice that the heat capacity is diverge at $r_\text{h} = L_\lambda$.  This indicates the distinction between two phases of black hole in asymptotically flat spacetime, namely the unstable small black hole ($r_\text{h}<L_\lambda$) with negative heat capacity ($C_\text{R}<0$) and the stable large black hole ($r_\text{h}>L_\lambda$) with positive heat capacity ($C_\text{R}>0$). Therefore the black holes can be in thermal equilibrium with infinite heat bath at a fixed temperature when its event horizon radius is larger than this characteristic length scale. This implies that the canonical ensemble exists for the Sch-flat from R\'enyi statistics.

The Hawking temperature of black hole is minimum at $r_\text{h}=L_\lambda$,  where it takes the form  
\begin{eqnarray}
T_{\text{min}}=\frac{1}{2}\sqrt{\frac{\lambda}{\pi}}=\frac{1}{2\pi L_\lambda}.
\end{eqnarray}
Below $T_{\text{min}}$, a black hole is not allowed to be created due to the absence of real value of $r_h$ as a solution of \eqref{TrS}.  Accordingly, this behavior of black holes in flat spacetime via R\'enyi statistics is in the same way as that of  AdS black holes in the GB statistics approach \cite{HawkingPage}. Interestingly, the presence of   $L_\lambda$ in the former case seems to be mathematically equivalent to that of  the AdS radius $L_{\text{AdS}}$ in the latter case. Moreover,  the existence of this characteristic length scale, in both cases, is a sign of the presence of small/large black hole phase transition.

\section{Generalized Smarr Formula} \label{Generalized Smarr Formula}

In the standard GB statistics, the first law of black hole thermodynamics can be written in the form
\begin{eqnarray}
dM = T_{\text{h}}dS_{\text{bh}} + \Omega dJ + \Phi dQ, \label{1stbht}
\end{eqnarray}
where $\Omega$ is the horizon angular velocity, $J$ is the angular momentum, $\Phi$ is the horizon electric potential, and $Q$ is the black hole's electric charge.  Applying the Euler's theorem for homogeneous function to the first law of black hole thermodynamics, the black hole mass can be written in the form~\cite{Smarr} 
\begin{eqnarray}
M = 2T_{\text{h}}S_{\text{bh}} + 2\Omega J + \Phi Q. \label{Smarr} 
\end{eqnarray}
Remark that this is the conventional Smarr formula within the framework of GB statistics.
On the other hand, we can obtain a modified version of  the Smarr formula from the R\'enyi statistics and treat the non-extensivity parameter $\lambda$ as a thermodynamic variable, which gives us the first law of black hole thermodynamics with the presence of the work term $PdV$. We will review shortly the process, as demonstrated in \cite{Chatchai2}, to obtain this generalized Smarr formula in the following. 

Considering a Schwarzschild black hole,  we assume that the nonextensive parameter is sufficiently small to avoid reaching the quantum gravity limit, as discussed in previous section.  Thus, we may use  $0<\lambda \ll 1$.  Substituting the Bekenstein-Hawking entropy $S_\text{bh}$ and Hawking temperature $T_\text{h}$ into \eqref{Srbh} and \eqref{T_R-T_Hawk}, we can obtain
\begin{eqnarray}
S_{\text{bh}} = \frac{e^{\lambda S_\text{R}}-1}{\lambda}\ \ \ \text{and} \ \ \ T_\text{h} = \frac{T_\text{R}}{e^{\lambda S_\text{R}}}. \label{S-T_bh_renyi}
\end{eqnarray}
We can reach a generalized Smarr formula within the R\'enyi statistics by putting \eqref{S-T_bh_renyi} into the Smarr formula  \eqref{Smarr} and expanding it in the power series of the parameter $\lambda$.   Then, we arrive at the formula
\begin{eqnarray}
M = 2T_\text{R}S_\text{R} - \lambda \frac{\pi r_\text{h}^3}{4} + \mathcal{O}(\lambda^2). \label{series}
\end{eqnarray}
 Here, we can introduce the thermodynamics pressure $P$ and the thermodynamic volume $V$ by identifying  
\begin{eqnarray}
P = \frac{3\lambda}{32} = \frac{3}{32\pi L_\lambda^2}, \ \ \ V = \frac{4}{3}\pi r_\text{h}^3. \label{extended} 
\end{eqnarray}
Keeping up to only the lowest order of $\lambda$, \eqref{series} becomes to be in a form of Smarr-like formula 
\begin{eqnarray}
M = 2T_\text{R}S_\text{R} - 2PV. \label{GenSmarr}
\end{eqnarray}
 Note that the dimension of $M$, $P$, $V$ in \eqref{extended} and \eqref{GenSmarr} coincides with the fact that $M\sim L$, $P\sim L^{-2}$ and $V\sim L^3$.  We can call this relation in \eqref{GenSmarr} as a generalized Smarr formula via R\'enyi statistics.
 
As shown above, $\lambda$ has been treated as a dynamical variable, namely it is associated with the thermodynamic pressure $P$. This notion and its corresponding first law of black hole thermodynamics has been proposed in \cite{Chatchai2}.  Following this approach, we substitute \eqref{S-T_bh_renyi} into \eqref{1stbht}, then the first law of black hole thermodynamics in the R\'enyi extended phase space of the Schwarzschild black hole with the variation of $\lambda$ takes the form
\begin{eqnarray}
dM &=& \frac{T_\text{R}}{e^{\lambda S_\text{R}}}d\left( \frac{e^{\lambda S_\text{R}}-1}{\lambda} \right), \nonumber \\
&=& T_\text{R}dS_\text{R}+\frac{1}{8}\pi r_\text{h}^3d\lambda + \mathcal{O}(\lambda^2).
\end{eqnarray}
Hence, keeping up to the first order in $\lambda$ and using the definition of $P$ and $V$ in \eqref{extended}, we arrive at the relation
\begin{eqnarray}
dM = T_\text{R}dS_\text{R} + VdP.
\end{eqnarray}
This result shows that the mass $M=M(S_\text{R}, P)$ is a function of $S_\text{R}$ and $P$.  This indicates that  the black hole mass $M$ is interpreted as the enthalpy $H(S_\text{R}, P)$ in the R\'enyi thermodynamics.   As the quantity conjugate to the pressure, the thermodynamic volume can be obtained from the relation $V = \left( \frac{\partial M}{\partial P} \right)_{S_\text{R}}$ at the first order of  $\lambda$ .  Through the Legendre transformation $U=H-PV$, the first law of black hole thermodynamics extends to be in the form
\begin{eqnarray}
dU = T_\text{R}dS_\text{R} - PdV. \label{ex1law}
\end{eqnarray}
In this framework, the extended first law of black hole thermodynamics due to the presence of the emergence $PdV$ term in \eqref{ex1law} becomes consistent with the first law of thermodynamics.  Since the non-extensive nature of black holes results from long-range gravitational interactions, the degree of freedoms of black holes do not only interact among themselves in the system but also interact with their environment. We have suggested in \cite{Chatchai2} that the correlations between black hole degrees of freedom and heat bath may be appeared as the additional work term, $PdV$, in \eqref{ex1law}. On the other hand, the emergence of this work term can occur as a consequence of the cosmological constant in the AdS black hole with the GB statistics in the framework of extended phase space~\cite{Kastor, Mann1, Kubiznak, Mann2, Mann3}.  Intriguingly, this is the equivalence between the non-extensivity parameter $\lambda$ in R\'enyi-flat black holes and the cosmological constant $\Lambda$ in AdS via GB-AdS black holes.   

\section{Solid/Liquid Phase Transition and Latent Heat via the R\'enyi Extended Phase Space} \label{Phase Transition}

To describe the thermal properties of gas, we use an equation of state to relate the thermodynamic quantities, such as pressure, volume and temperature in a single equation. The extended first law of black hole thermodynamics \eqref{ex1law} is very similar to a system of gas enclosed in the volume $V$ and subject to an external pressure $P$. From this similarity, we will construct an equation of state to explore thermodynamic properties and phase structure of black hole by substituting the non-extensivity parameter $\lambda =\frac{32P}{3}$ and the event horizon radius $r_\text{h}=\left(\frac{3V}{4\pi}\right)^{\frac{1}{3}}$ into the temperature expression \eqref{TrS}. Solving $P=P(T_\text{R}, V)$, we obtain an equation of state of the Sch-flat black hole as

\begin{figure}
	\centering
	\includegraphics[scale=0.45]{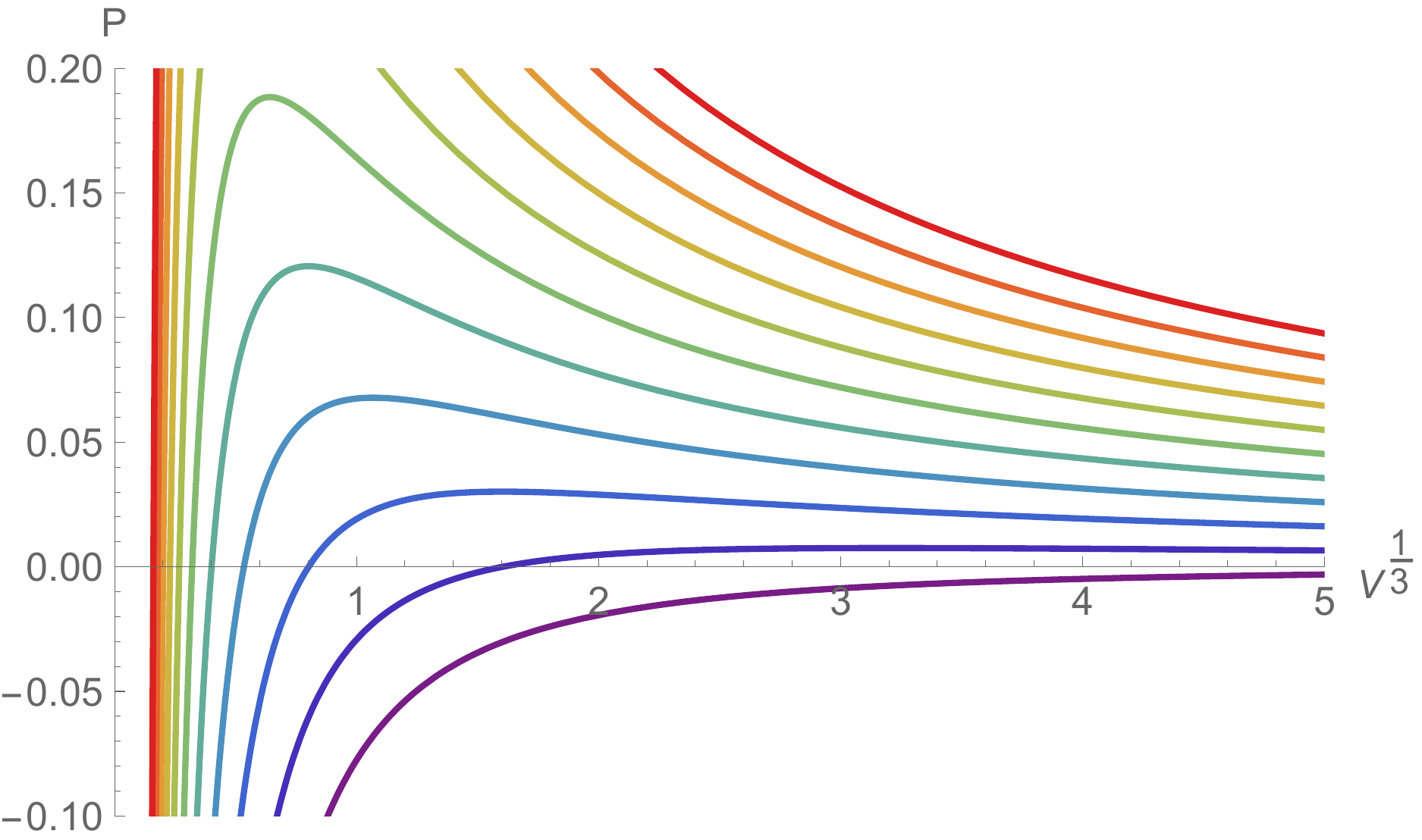}
	\caption{The isothermal curves in the $P-V^{1/3}$ diagram of the Sch-flat in R\'enyi extended phase space is shown. The temperature decreases from top to bottom. At a given temperature, there are two phases of black holes, namely the small black hole and large black hole which correspond to positive and negative slope in $P-V^{1/3}$ plane restpectively. The shape of isothermal curve in the $P-V$ plane is the same but more horizontally stretched.}\label{fig:5}
\end{figure}

\begin{eqnarray}
P = \frac{1}{8}\left( \frac{4\pi}{3} \right)^{\frac{1}{3}}\left( \frac{3T_\text{R}}{V^{\frac{1}{3}}} - \left(\frac{3}{4\pi}\right)^{\frac{2}{3}}\frac{1}{V^{\frac{2}{3}}} \right). \label{eos}
\end{eqnarray}
For this equation of state, the curves of fixed temperature are plotted in \figref{fig:5}, where the pressure has a maximum value $P_\text{max}$ at volume $V_o$ as
\begin{eqnarray}
P_\text{max}=\frac{3\pi}{8}T^2_\text{R}\ , \ \ \ V_o = \frac{1}{(6\pi^2)^{\frac{1}{3}}T_\text{R}}.
\end{eqnarray}
Below $P_\text{max}$, there are two black hole phases associate to small and large black hole configuration when $V<V_o$ and $V>V_o$, respectively. The R\'enyi entropy given in term of $P$ and $V$ is
\begin{eqnarray}
S_\text{R}(P, V) = \frac{3}{32P}\ln \left( 1+16\left( \frac{\pi}{6} \right)^{\frac{1}{3}}PV^{\frac{2}{3}} \right). \label{Spv}
\end{eqnarray}

\begin{figure*}[!ht]
	\begin{tabular}{c c}
		\includegraphics[scale=0.4]{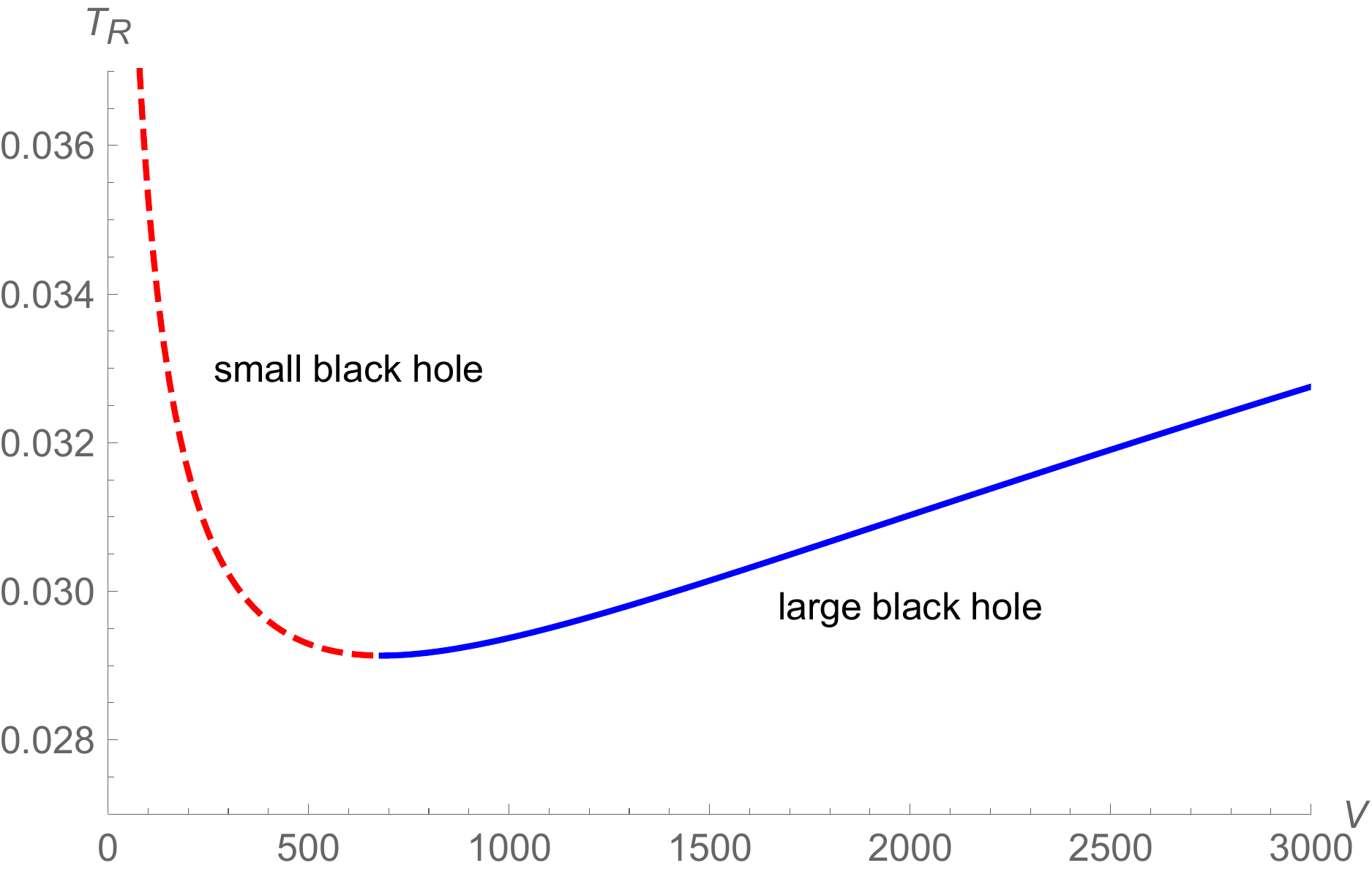}\quad
		\includegraphics[scale=0.39]{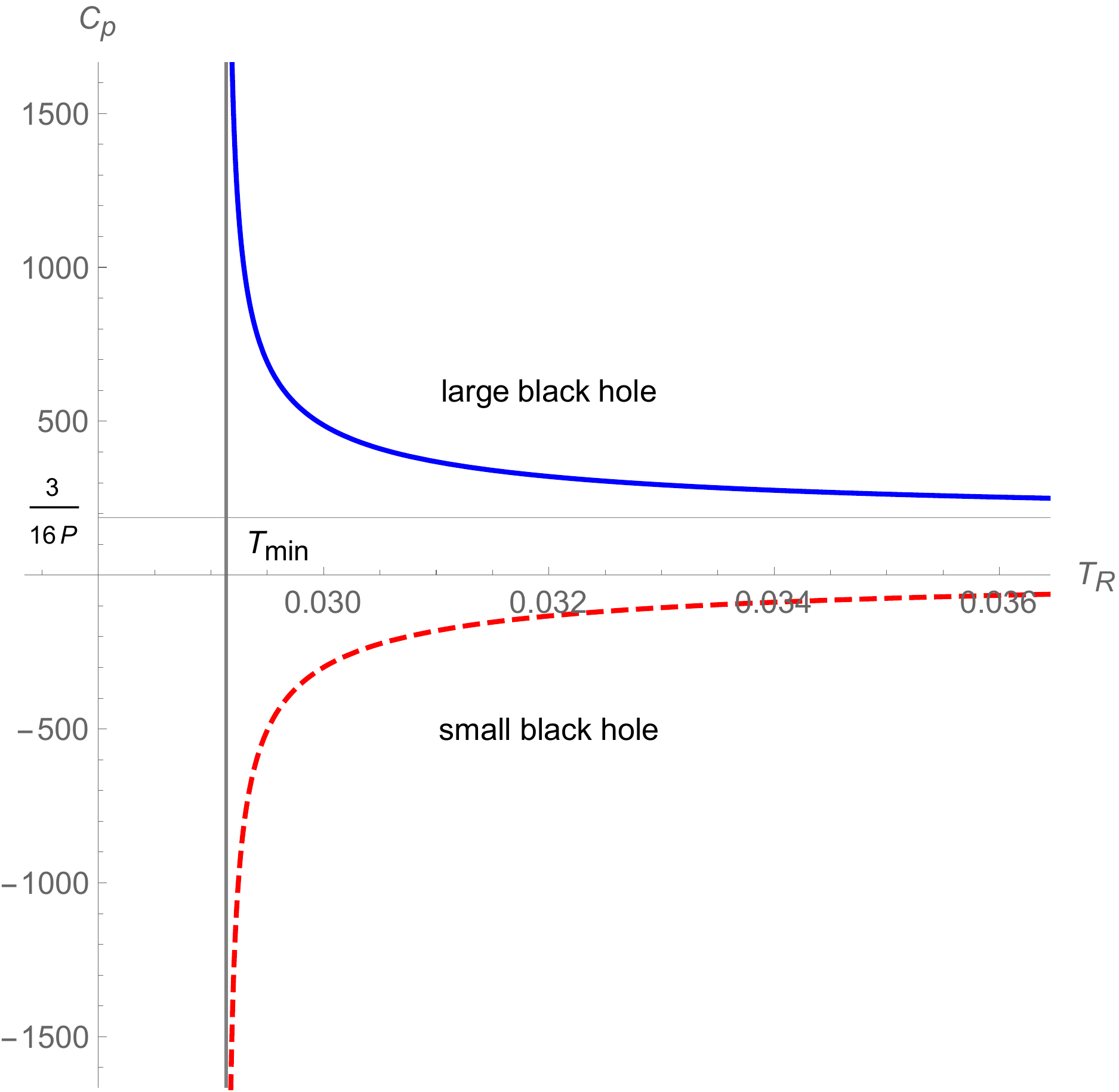}
	\end{tabular}
	\caption{Left: The R\'enyi temperature of a Schwarzschild black hole $T_\text{R}$ versus the thermodynamic volume $V$ is plotted with $P=$ 0.001. Right: The coresponding heat capacity at constant pressure $C_\text{P}$ of a black hole versus $T_\text{R}$. In both panels, the dashed red and solid blue lines represent small and large black hole, restpectively. Black hole solution do not exist when $T<T_{\text{min}}$.}\label{fig:2}
\end{figure*}

The heat capacity at constant pressure can be expressed in the form
\begin{eqnarray}
C_\text{P} = T_\text{R}\left( \frac{\partial S_\text{R}}{\partial T_\text{R}} \right)_P = -3\left(\frac{\pi}{6}\right)^{\frac{1}{3}}\frac{V^{\frac{2}{3}}}{1-\left( \frac{V}{V_c} \right)^{\frac{2}{3}}},
\end{eqnarray}
where $V_c=\frac{1}{32}\sqrt{\frac{3}{2\pi}}\frac{1}{P^{\frac{3}{2}}}$. Obviously, the heat capacity is negative when $V<V_c$, positive when $V>V_c$ and diverge at $V=V_c$. In the large volume limit $V\rightarrow \infty$, the heat capacity converge to the positive finite value $C_\text{P} = \frac{3}{16P}$. The temperature is minimum at the volume $V_c$, which can be described in term of the pressure $P$ as
\begin{eqnarray}
T_{\text{min}} = \sqrt{\frac{8P}{3\pi}}.
\end{eqnarray}
For $T<T_{\text{min}}$, there is no black hole solution, the system is in the thermal radiation phase. This is illustrated in \figref{fig:2}.

Thermal fluctuations can induce the microscopic degree of freedom rearrange, resulting in a phase transition of macroscopic state of matter. This interesting phenomena also occur in gravitational physics, such as the nucleation of AdS black hole in a thermal heat bath. The thermodynamic stability of thermal phase under constant pressure and temperature can be described by Gibbs free energy. In a thermal gravitational system, the Gibbs free energy can be calculated from the standard Euclidean action method. Since, we used R\'enyi statistics to decribes the thermodynamics of black hole, the on-shell Gibbs free energy should be replaced by
\begin{figure*}[!ht]
	\begin{tabular}{c c}
		\includegraphics[scale=0.37]{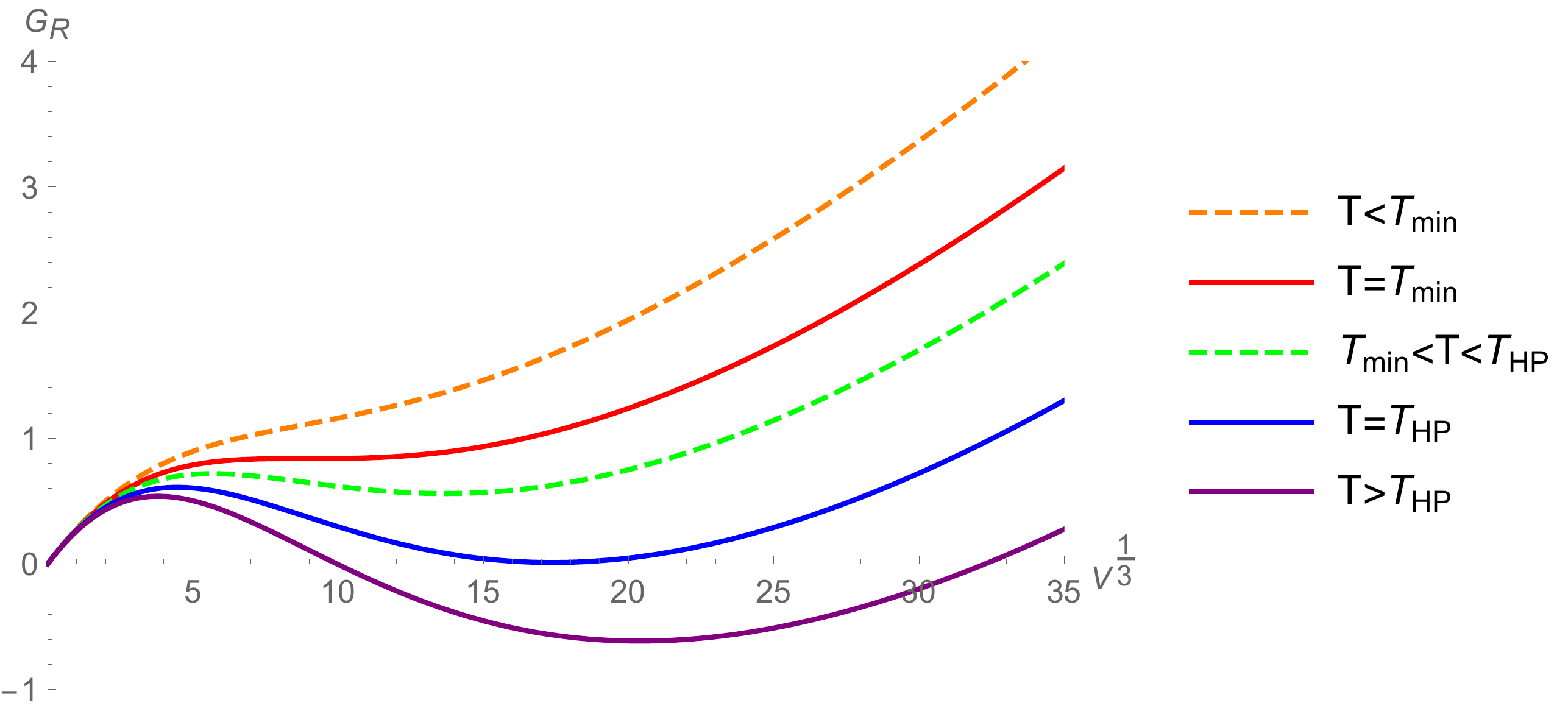}\quad
		\includegraphics[scale=0.34]{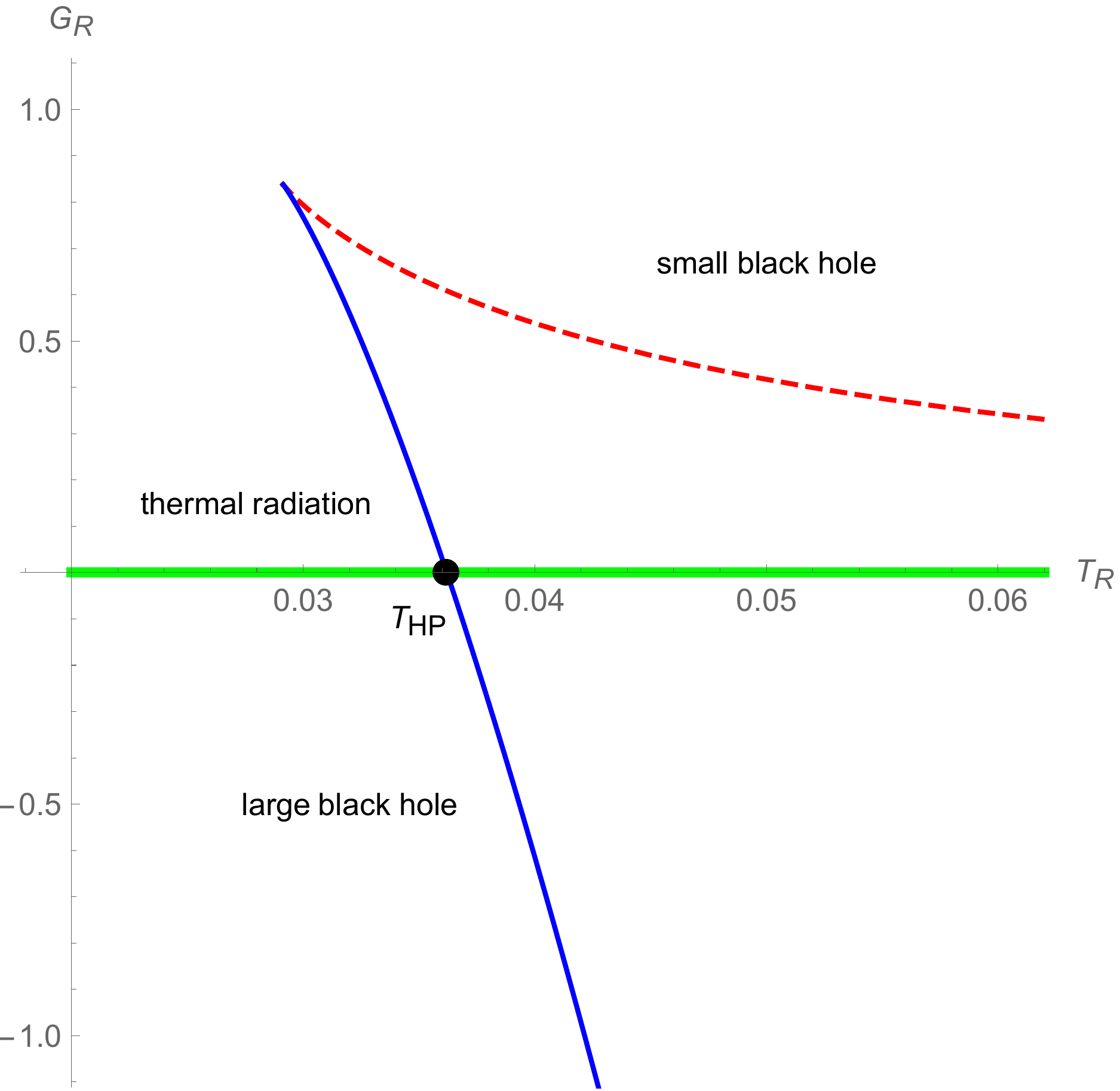}
	\end{tabular}
	\caption{Left: The off-shell Gibbs free energy of a Schwarzschild black hole $G_\text{R}$ versus the thermodynamic volume $V^{\frac{1}{3}}$ is plotted with $P=$ 0.001. Note that, the curve of $G_\text{R}$ versus $V$ looks similiar to this graph but its horizontal coordinate is much more extended. Right: The on-shell Gibbs free energy $G_\text{R}$ as a function of R\'enyi temperature $T_\text{R}$ is shown.}\label{fig:3}
\end{figure*}
\begin{eqnarray}
G_\text{R} = M-T_\text{R}S_\text{R}, \label{free}
\end{eqnarray}
where $M$ is the enthalpy of the system. As mentioned in section \ref{R-stat&BH thermo}, the canonical ensemble with fixed temperature exists in the consideration with R\'enyi statistics.  Following from \cite{Spallucci, Li}, we can construct the off-shell Gibbs free energy of the black hole in the canonical ensemble at fix temperature $T$ by substituting the black hole's temperature $T_\text{R}$ in \eqref{free} with the ensemble temperature $T$. In this way, we obtain  
\begin{eqnarray}
G_\text{R} = \frac{1}{2}\left( \frac{3}{4\pi} \right)^{\frac{1}{3}}V^{\frac{1}{3}}-\frac{3}{32}\frac{T}{P}\ln \left[ 1+16\left( \frac{\pi}{6} \right)^{\frac{1}{3}}PV^{\frac{2}{3}}\right].
\end{eqnarray}
Here, we have identified the thermodynamic volume $V$ as an order parameter. To obtain the black hole configuration in the heat bath, we replace the black hole's temperature $T_\text{R}$ in \eqref{eos} by the ensemble temperature $T$, then the order parameter $V$ of the small and large black hole can be solved analytically by
\begin{eqnarray}
V^{\frac{1}{3}}_{\text{LBH/SBH}}=\left( \frac{4\pi}{3}\right)^{\frac{1}{3}}\frac{T}{2\pi T^2_{\text{min}}}\left( 1 \pm \sqrt{1-\frac{T^2_{\text{min}}}{T^2}} \right), \label{slv}
\end{eqnarray}  
where $V_{\text{LBH}}$ and $V_{\text{SBH}}$ represent the volume of large and small black hole, respectively. In the configuration that there is no black hole, namely in the case $r_\text{h}=0$,  a Minkowski spacetime filled with thermal radiation represents classical hot flat space phase, which has the order parameter $V_{\text{rad}}=0$.   Moreover, the Gibbs free energy and the entropy of this phase are virtually zero because the number of quantum particles associated to the radiation occupying the spacetime is very small.

The plots of the off-shell and on-shell Gibbs free energy are shown in \figref{fig:3}, the Hawking-Page like phase transition emerge in the similar way as the black hole in AdS space via GB statistics \cite{HawkingPage}. In a very low temperature limit $T<T_{\text{min}}$, there is no black hole exist since the order parameter of black hole in \eqref{slv} is an imaginary number. The only one global minimum of the $G_\text{R}$ curve is at $V_{\text{rad}}=0$, representing a pure thermal radiation in an asymptotically flat. Interestingly, ours result is in contrast from the GB statistics, which show that the black hole has no minimum bound on its temperature. If we increase the temperature until $T=T_{\text{min}}$, the black hole nucleation start here at an inflection point $V_c$ of $G_\text{R}$. 

For $T_{\text{min}}<T<T_{\text{HP}}$,  the small and large black hole phases emerged corespond to a local maximum and a local minimum of $G_\text{R}$ curve, respectively. However, the two locally extremum has higher free energy than the thermal radiation phase therefore the most thermodynamical prefer phase is still the pure thermal radiation state.

\begin{figure}
	\centering
	\includegraphics[scale=0.38]{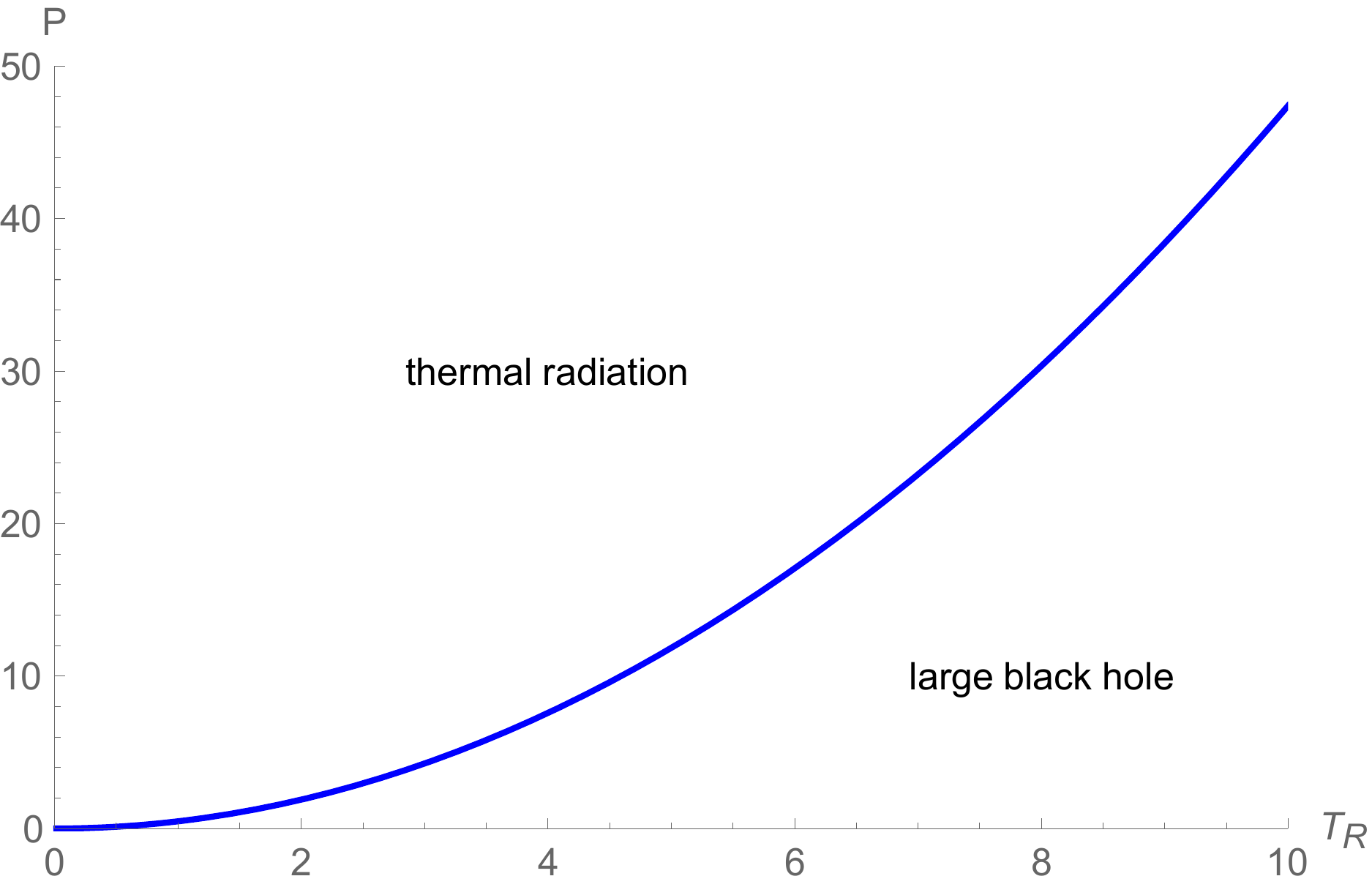}
	\caption{The $P-T_\text{R}$ phase diagram of the first order, the so-called Hawking-Page phase transition, between thermal radiation and large black hole via R\'enyi statistics. Note that the coexistence line continues extending and not stop running at high pressure and temperature. This implies the absence of a critical point, and then support that this phase transition is similar to a solid/liquid phase transition, instead of liquid/gas phase transition.}\label{fig:4}
\end{figure}

At $T=T_{\text{HP}}$, the two local minimum are degenerate, namely the thermal radiation phase and the large black hole phase, suggesting a coexistence phase between them. The Hawking-Page temperature is determined from $G_\text{R}=0$, which can be described numerically by
\begin{eqnarray}
T_{\text{HP}}  \sim 1.14\sqrt{P}. \label{Thp}
\end{eqnarray}
The coexistence line of thermal radiation/large black hole phases can be written in the form
\begin{eqnarray}
P\mid _\text{coexistence}\sim 0.769 \ T^2_\text{R}.
\end{eqnarray}
The coexistence line in the $P-T_\text{R}$ plane is plotted in \figref{fig:4} ~has no terminate point because $P$ in \eqref{Thp} has no bound, so the Hawking-Page phase transition can occur at all pressure. This reminiscent of a solid/liquid phase transition \cite{Kubiznak, Mann2}. When the system cross this coexistence line, the order parameter $V_{\text{rad}}=0$ of thermal radiation phase jump discontinuously to large black hole phase 

\begin{eqnarray}
V_{\text{HP}}\sim \frac{0.168}{P^{\frac{3}{2}}}, \label{vhp} 
\end{eqnarray}
in the $G_\text{R}$ curve, which is the characteristic of the first order phase transition. In this way, the entropy of the system \eqref{Spv} also change discontinuously, where it takes the form
\begin{eqnarray}
\Delta S = S_\text{LBH} - S_\text{rad} = \frac{3}{32P}\ln \left( 1+16\left( \frac{\pi}{6} \right)^{\frac{1}{3}}PV_{\text{HP}}^{\frac{2}{3}} \right), \label{ds}
\end{eqnarray}
where $S_\text{rad}\sim 0$. At the coexistence line, their Gibbs free energies are equal, {\it i.e.} $G_\text{rad}(T_\text{HP})=G_\text{LBH}(T_\text{HP})$, thus we obtain 
\begin{eqnarray}
E_\text{rad}-T_\text{HP}S_\text{rad}&=&M_\text{LBH}-T_\text{HP}S_\text{LBH}, \nonumber 
\end{eqnarray}
where $E_\text{rad}$ is the energy of thermal radiations.  From this, we then obtain
\begin{eqnarray}
S_\text{LBH}-S_\text{rad}&=&\frac{M_\text{LBH}-E_\text{rad}}{T_\text{HP}} \nonumber \\
\Delta S&=& \frac{L}{T_\text{HP}},
\end{eqnarray} 
where we have replaced $M_\text{LBH}-E_\text{rad}$ by the latent heat $L$ of the transition to nucleate the large black hole in flat spacetime from the pure thermal radiations, and vice versa.  This can be done because the latent heat is usually defined by 
\begin{eqnarray}
L = T_{\text{HP}}\Delta S.   \label{latent}
\end{eqnarray}
By substititing \eqref{Thp}, \eqref{vhp} and \eqref{ds} into \eqref{latent}, we obtain
\begin{eqnarray}
L \sim \frac{0.171}{\sqrt{P}}.
\end{eqnarray}
Notice that it is equal to the black hole's mass on the coexistence line $M_{\text{HP}}= \frac{1}{2}\left( \frac{3V_{\text{HP}}}{4\pi} \right)^{\frac{1}{3}}$. Ours result is very similar to the black hole in AdS from GB statistics \cite{Belhaj, Dolan}. In a case of the Sch-flat with GB statistics, the pressure $P=0$, the latent heat tend to infinite which mean that the thermal radiation cannot classically nucleate to form a black hole in asymptotically flat spacetime. Finally, at the high temperature limit $T>T_{\text{HP}}$, the large black hole phase is the most thermodynamically prefer state since its $G_\text{R}$ becomes the absolute globally minimum value. 

\section{Black Hole Heat Engine} \label{BH heat engine}

We have already discussed the generalized Smarr formula via R\'enyi statistics in section \ref{Generalized Smarr Formula} which allows the work term emerges in the first law of black hole thermodynamics. In this section, we will investigate further the second law of black hole thermodynamics within the R\'enyi extended phase space approach. To achieve this, we study the Carnot heat engine by using the black hole as a working substance.  As stated in the Carnot theorem, every reversible heat enginie perfoming between hot and cold energy reservoirs have the same efficiency, regardless of the type of engine and working substance. One question may be arisen whether, within the R\'enyi extended thermodynamics,  the black hole heat engine  can give the Carnot efficiency.  

Let us start with reviewing some theoretical backgrounds about heat engine.
Generically, a cyclic thermodynamic process operates between the hot and the cold heat reservior with the temperature $T_\text{H}$ and $T_\text{C}$, respectively.  In each cycle, it performs the net work  $W_{\text{eng}}$, whereas it absorbs the heat from the hot reservoir $|Q_\text{H}|$ and expels the heat to the cold reservior $|Q_\text{C}|$.  From the first law of thermodynamics, the thermal efficiency of a heat engine is in the form
\begin{eqnarray}
\eta = \frac{W_{\text{eng}}}{|Q_\text{H}|} = 1-\frac{|Q_\text{C}|}{|Q_\text{H}|}, \label{eff}
\end{eqnarray}
which is always less than one, according to the second law of thermodynamics.  However, the theoretical maximum efficiency, namely the Carnot effieciency, can be obtained by applying the second law of thermodynamics. We will demonstrate it here.   

Consider the total change of entropy, including the system and environment.  Through each cycle, the total entropy is  $\Delta S_{\text{total}} = \Delta S_{\text{eng}}+\Delta S_\text{env}$.  The change of environment consists of the hot and cold heat reservoirs. Therefore, the change of environment entropy is the sum of the change of the entropy of hot and cold heat reservoirs, that is  $\Delta S_\text{H}$  and  $\Delta S_\text{C}$, respectively.   Since the cyclic process here is the reversible process, we have $\Delta S_{\text{eng}}=0$ after each cycle finish. The total entropy $\Delta S_\text{total}$ is reduced to be only in the term of $\Delta S_\text{env}$.  Hence, the second law reads
\begin{eqnarray}
	\Delta S_{\text{total}} =\Delta S_\text{H}+\Delta S_\text{C} \geq 0. \label{gen2nd}
\end{eqnarray}
From the formula $\Delta S = \frac{\Delta Q}{T}$, the entropy change of the hot and cold heat reservior are $\Delta S_\text{H}=-\frac{|Q_\text{H}|}{T_\text{H}}$ and $\Delta S_\text{C}=\frac{|Q_\text{C}|}{T_\text{C}}$ restpectively. Substituting them into \eqref{gen2nd}, we have 
\begin{eqnarray}
-\frac{|Q_\text{H}|}{T_\text{H}}+\frac{|Q_\text{C}|}{T_\text{C}} & \geq 0& ,\nonumber \\
-\frac{|Q_\text{H}|}{T_\text{H}}+\frac{|Q_\text{H}|-W_{\text{bh}}}{T_\text{C}} & \geq 0& ,\nonumber \\
\frac{W_{\text{bh}}}{|Q_\text{H}|} &\leq & 1-\frac{T_\text{C}}{T_\text{H}} \nonumber \\
\eta &\leq & 1-\frac{T_\text{C}}{T_\text{H}} .
\end{eqnarray}
Therefore, the maximum value of $\eta$ is $\eta_\text{C}=1-\frac{T_\text{C}}{T_\text{H}}$. The $\eta_\text{C}$ is the efficiency of heat engine operates along the Carnot cycle. Carnot's theorem can be stated that ``No engine operating between two heat reserviors has more efficient than the Carnot engine''. Obviously, the two statements, Carnot's theorem and the GSL are equivalence.

Here, the black hole can be in thermal equilibrium with an infinite heat reservior. Basically, this is a condition necessary for the existence of a quasi-static change in the reversible process and the emergence of mechanical work from heat via the $PdV$ tem in \eqref{ex1law}. We can construct the heat engine operating through the Carnot cycle in the context of asymptotically flat black hole.  Before mentioning the Carnot cycle of black hole, it is important to derive some useful relations of the internal energy from the eqation of state \eqref{eos}. Let us consider the first law of black hole thermodynamics \eqref{ex1law}   
\begin{eqnarray}
T_\text{R}dS_\text{R} = dU + PdV. \label{1st}
\end{eqnarray}
The R\'enyi entropy can be written in the form $S_\text{R} = S_\text{R}(T_\text{R}, V)$, therefore its differntial form is given by
\begin{eqnarray}
dS_\text{R} = \left( \frac{\partial S_\text{R}}{\partial T_\text{R}} \right)_V dT_\text{R} + \left( \frac{\partial S_\text{R}}{\partial V} \right)_{T_\text{R}} dV.
\end{eqnarray}
Multiplying both sides of the above equation by $T_\text{R}$ and using the Maxwell relation, $\left( \frac{\partial S_\text{R}}{\partial V} \right)_{T_\text{R}} = \left( \frac{\partial P}{\partial T_\text{R}} \right)_V$, we obtain
\begin{eqnarray}
T_\text{R}dS_\text{R} = C_\text{V}dT_\text{R} + T_\text{R}\left( \frac{\partial P}{\partial T_\text{R}} \right)_VdV, \label{2nd}
\end{eqnarray}
where $C_\text{V} = T_\text{R}\left( \frac{\partial S_\text{R}}{\partial T_\text{R}} \right)_V$ that is the heat capacity at fixed constant volume. Combining \eqref{1st} and \eqref{2nd}, the variation of internal energy becomes
\begin{eqnarray}
dU = C_\text{V} dT_\text{R} + \left[ T_\text{R}\left( \frac{\partial P}{\partial T_\text{R}} \right)_V - P \right]dV. \label{du}
\end{eqnarray}
Because $dU$ is an exact differential equation, we have
\begin{eqnarray}
\frac{\partial C_\text{V}}{\partial V} = \frac{\partial}{\partial T_\text{R}}\left[ T_\text{R}\left( \frac{\partial P}{\partial T_\text{R}} \right)_V - P \right] =T_\text{R} \left( \frac{\partial^2 P}{\partial T_\text{R}^2} \right)_V.
\end{eqnarray}
From the equation of state in \eqref{eos}, we found that $\left( \frac{\partial^2 P}{\partial T_\text{R}^2} \right)_V=0$, this mean that $C_\text{V}$ does not depend on the thermodynamics volume $V$ and depend only on the temperature $T_\text{R}$ \cite{Agrawal, Santos, Tjiang}. Substituting $dU$ from \eqref{du} into \eqref{1st}, 
\begin{eqnarray}
dQ &=& dU + PdV, \nonumber \\
&=& C_\text{V}dT_\text{R} + T_\text{R}\left( \frac{\partial P}{\partial T_\text{R}} \right)_V dV, \nonumber \\
&=& C_\text{V}dT_\text{R} + \frac{3}{8}\left( \frac{4\pi}{3} \right)^{\frac{1}{3}}\frac{T_\text{R}}{V^{\frac{1}{3}}}dV,   \label{dq}
\end{eqnarray}
where $dQ=T_\text{R}dS_\text{R}$ . Note that, we used an equation of state \eqref{eos} to obtain the second term of \eqref{dq}.

We consider the black hole as a working substance that operates between hot and cold reserviors with the temperature $T_\text{H}$ and $T_\text{C}$, respectively. One of the basic Carnot cycle consists of the four steps as shown in \figref{fig:5}.  In the isothermal process $A\rightarrow B$, the heat engine performs the mechanical work the isothermal expansion whereas it absorbs the heat from energy reservior at the temperature $T_\text{H}$.  The relation between the absorbed heat $ Q_\text{H}$, $T_\text{H}$ and the volume $V_\text{A}$ and $V_\text{B}$ can be obtained by integrating \eqref{dq} as follows
\begin{eqnarray}
Q_\text{H} &=& \int_A^B dQ \nonumber \\ 
&=& \frac{3}{8}\left( \frac{4\pi}{3} \right)^{\frac{1}{3}}T_\text{H} \int_{V_\text{A}}^{V_\text{B}} \frac{dV}{V^{\frac{1}{3}}} \nonumber \\
&=& \frac{9}{16}\left( \frac{4\pi}{3} \right)^{\frac{1}{3}}T_\text{H} (V_\text{B}^{\frac{2}{3}}-V_\text{A}^{\frac{2}{3}}), \label{QH}
\end{eqnarray}
where the first term of \eqref{dq} is zero since $dT_\text{R}=0$ along the isothermal path.   In a similar way, the isothermal compression process $C\rightarrow D$ in the contact with the cold energy reservior of the temperture $T_\text{C}$ exhausts the heat of the form
\begin{eqnarray}
Q_\text{C} &=& \int_C^D dQ \\ \nonumber
                 &=&  - \frac{9}{16}\left( \frac{4\pi}{3} \right)^{\frac{1}{3}}T_\text{C} (V_\text{C}^{\frac{2}{3}}-V_\text{D}^{\frac{2}{3}}). \label{QC}
\end{eqnarray}
Manifestly, since $V_\text{B}>V_\text{A}$ and $V_\text{C}>V_\text{D}$, it can be seen from \ref{QH} and \ref{QC} that the heat $Q_\text{H}>0$ and $Q_\text{C}<0$.

On the other hand, the path $B\rightarrow C$ and $D\rightarrow A$ are adiabatic process, along which $dQ=0$. Integrating \eqref{dq} for these two adiabatic processes, we have 
\begin{eqnarray}
\int_{T_\text{R,B}}^{T_\text{R,C}} \frac{C_\text{V}}{T_\text{R}}dT_\text{R} + \frac{9}{16}\left( \frac{4\pi}{3} \right)^{\frac{1}{3}} (V_\text{C}^{\frac{2}{3}}-V_\text{B}^{\frac{2}{3}}) = 0, \label{adia1}
\end{eqnarray}
and
\begin{eqnarray}
\int_{T_\text{R,D}}^{T_\text{R,A}} \frac{C_\text{V}}{T_\text{R}}dT_\text{R} + \frac{9}{16}\left( \frac{4\pi}{3} \right)^{\frac{1}{3}} (V_\text{A}^{\frac{2}{3}}-V_\text{D}^{\frac{2}{3}}) = 0. \label{adia2}
\end{eqnarray}
Combining \eqref{adia1} and \eqref{adia2}, we obtain the condition
\begin{eqnarray}
V_C^{\frac{2}{3}}-V_B^{\frac{2}{3}} = -(V_A^{\frac{2}{3}}-V_D^{\frac{2}{3}}), \ \ \ \text{or} \ \ \ V_C^{\frac{2}{3}}-V_D^{\frac{2}{3}} = V_B^{\frac{2}{3}}-V_A^{\frac{2}{3}}. \label{adiabat}
\end{eqnarray}
Since $\frac{C_\text{V}}{T_\text{R}}$  depends only on the temperature and $T_\text{R,A}=T_\text{R,B}=T_\text{H}$ and $T_\text{R,C}=T_\text{R,D}=T_\text{C}$, we arrive at the formula
\begin{eqnarray}
\int_{T_\text{R,B}}^{T_\text{R,C}} \frac{C_\text{V}}{T_\text{R}}dT_\text{R} + \int_{T_\text{R,D}}^{T_\text{R,A}} \frac{C_\text{V}}{T_\text{R}}dT_\text{R}=0.
\end{eqnarray}
The thermal efficiency of the Carnot engine using the Sch-flat black hole as a working substtance $\eta_c$ can be obtained by substituting \eqref{QH} and \eqref{QC} into \eqref{eff} and using the relation \eqref{adiabat}.  Hence, we obtain the thermal efficiency
\begin{eqnarray}
\eta_c = 1 - \frac{|Q_\text{C}|}{|Q_\text{H}|} = 1 - \frac{T_\text{C}}{T_\text{H}}.
\end{eqnarray}
Interestingly, the equation of state of the Sch-flat black hole within the R\'enyi extended phase space approach gives rise to the thermal efficiency in the reversible process $\eta_c$ following the Carnot's theorem in the same way as occuring in the conventional matter. This implies that no black hole heat engine can convert heat from a hot reservior into mechanical work without some energy released to the environment.  Therefore, the R\'enyi extended phase space approach in black hole thermodynamics renders the thermodynamic description satisfying the second law of thermodynamics or the GSL.  Remarkably, our results have given a further evidence that treating the nonextensivity parameter $\lambda$ as the thermodynamic pressure $P$, as proposed in \cite{Chatchai2}, surprisingly brings about a consistent thermodynamics.

\section{Conclusion and Disscussion} \label{Conclusion}

After reviewing the nonextensive entropy approach in black hole thermodynamics in section~\ref{Tsallis-R stat}, \ref{R-stat&BH thermo} and \ref{Generalized Smarr Formula}, we explore the R\'enyi thermodynamics of the Schwarzschild black hole in asymptotically flat spacetime in section~\ref{Phase Transition} as the first part of our study.   The results indicate that, with $0<\lambda <1$, there are small and large black hole phases emerging while this cannot occur in the GB statistics. Having introduced $L_\lambda = 1/\sqrt{\pi \lambda}$ as the nonextensivity length scale in \cite{Chatchai2}, we find here that as long as the black hole has the size $r_h$ greater than this length scale, it has positive heat capacity. Namely, the large black hole can be in thermal equilibrium with thermal radiation at a fixed temperature.  This implies that the canonical ensemble can exist in the considertion with this alternative R\'enyi entropy.

Treating the nonextensivity parameter $\lambda$ as a thermodynamic variable, it can be assigned as a thermodynamic pressure whereas its conjugate variable is a thermodynamic volume. In this extended approach, the mass of black hole represents the enthalpy rather than the internal energy. Interestingly, the mechanical work $PdV$ term exists in the first law of black hole thermodynamics \eqref{ex1law}, which allows us to construct the black hole heat engine, as discussed in this paper.  We write the equation of state in term of pressure $P$ and volume $V$, and study the thermal phase structure of the Sch-flat black hole in this R\'enyi extended phase space approach.  The black hole exhibits the first order Hawking-Page phase transition between thermal radiation phase and large black hole phase which can be determined from the off-shell Gibbs free energy. The coexistence curve between thermal radiation and large black hole in the $P-T$ plane is equivalent to the semi-infinite quadratic curve, which is reminiscent of a solid/liquid phase transition.  The latent heat of fusion from solid (corresponding to thermal radiation) to liquid (corresponding to large black hole) is inversely proportional to the square root of pressure, \textit{i.e.} $L\sim 1/\sqrt{P}$.  We can deduce from this that the phase of thermal radiations cannot classically turn into the situation of a black hole nucleation within the GB statistics where $P=0$.

In the second part of our study in this paper, {\it i.e.} section~\ref{BH heat engine}, we investigate the validity of the GSL in the context of R\'enyi extended phase space approach.  Interestingly, the quasi-static change of the black hole thermodynamical system in the reversible process is possible in this approach because the black hole can be in thermal equilibrium with its surrounding and there exists the emerging mechanical work from heat via the $PdV$ term.  Consequently, we can construct a heat engine by taking the Sch-flat black hole as a working substance.  We define a Carnot cycle in the $P-V$ plane and derive the thermal efficiency of Sch-flat black hole heat engine. Intriguingly, we obtain the efficiency in the form $\eta_c=1-T_\text{C}/T_\text{H}$. This maximum possible efficiency of reversible heat engine follows the second law since it gives the result corresponding to the Carnot theorem in the usual GB thermodynamics. In other words, the second law of thermodynamics or GSL is still satisfied even in the alternative R\'enyi thermodynamics.  This gives more evidence that interpreting the parameter $\lambda$ as thermodynamic pressure and treaing its conjugate as volume could provide a proper thermodynamic description.

In this work, our motivation is to demonstrate that the R\'enyi extended phase space approach may be a reasonable statistical model of black holes by considering some macroscopic phenomena including the thermal phase transition and the reversible heat engine.  These are some embodiments that our consideration is in a well-behaved thermodynamic description, especially satisfying the second law of thermodynamics.   However, we did not so far discuss the microscopic description or  quantum originated aspects of the R\'enyi statistics approach to black holes thermodynamics.  Let us spend the space here to discuss some potentially physical implications of the R\'enyi model in describing the quantum aspects of black holes.

Generically, in the long-range interaction systems like a self-gravitating system, their microstates could be correlated in some way with the heat reservior.  As we suggested in \cite{Chatchai2}, some energy density emerging from the nontrivial correlation between black hole and heat reservior may be encoded in the nonextensive parameter $\lambda$.  This emergent energy density from the correlations induces a pressure to the black holes in the same way as the cosmological constant in the case of AdS black holes through the extended phase space approach.  It is important to emphasize that the appearance of this energy density in black hole thermodynamics is an emergent phenomena whereas there is no its contribution explicitly in the geometry of classical black hole solution. Therefore, there is no directly mathematical parallel between geometric and thermodynamic descriptions in black hole thermodynamics from the R\'enyi statistics, which is defferent from the results of the consideration using the GB statistics.

To support the idea that the modified thermodynamics from R\'enyi statistics results from the correlations between a black hole and its environment, let us consider the scaling of the black hole's R\'enyi entropy, as shown in \eqref{SrS}, with the horizon radius in the limit of small and large black holes, respectively, as follows:
\begin{equation}
S_\text{R} = \begin{cases*}
  \pi r^2_\text{h}, & if $r_\text{h}\ll L_\lambda$,\\
  \frac{2}{\lambda}\ln \left( \frac{r_\text{h}}{L_\lambda}  \right),              & if $r_\text{h}\gg L_\lambda$. \label{scaling}  
\end{cases*}
\end{equation}
In the $r_\text{h}\ll L_\lambda$ (small black hole) limit, it is obvious that the R\'enyi entropy approximately obeys an area law as the Bekenstein-Hawking entropy in the GB statistics.  The area scaling of black hole's entropy in this $\lambda\to 0$ limit may imply that the entropy corresponds to an entanglement entropy~\cite{Bombelli, Srednicki, Eisert}.  What about the logarithmic scaling of the black hole entropy in the $r_\text{h}\gg L_\lambda$ (large black hole) limit?  Apparently, there is a violation of the area law when the horizon radius is much larger than the nonextensivity length scale. This logarithmic scaling entropy is reminiscent of some lattice models in condensed matter physics, which includes for example one-dimensional quantum spin chain just away from criticality with the entanglement entropy of the form~\cite{Vidal, Cardy}
\begin{equation}
S_\text{A}=\frac{c}{6}\mathcal{A}\ln \frac{\xi}{a}. \label{ee}
\end{equation}
Note that $\xi$ is a correlation length, $a$ is a lattice spacing and $\mathcal{A}$ is the number of boundary points between the system A and its complement B.   The expression above can be obtained from the von Neumann entropy of the subsystem A, where the degrees of freedom of the inaccessible subsystem B is traced out.  Comparing this system with the Sch-flat one, the black hole analogue of B could be the interior region covered by the event horizon in which the information cannot be reachable. On the other hand, the subsystem A may correspond to an exterior region outside the event horizon where an observer is sitting. Matching \eqref{scaling} in the large black hole limit with \eqref{ee}, we may identify that $r_\text{h}$ and $L_\lambda$ are associated with the correlation length $\xi$ and the lattice spacing $a$, respectively.  Thus, these could be written symbolically as
\begin{equation}
\begin{array}{l}\nonumber
	r_h \longleftrightarrow \xi ,\\ 
	L_{\lambda} \longleftrightarrow a,
\end{array}
\end{equation}
where $\lambda$ may be related to the central charge $c$ in the form of $\lambda =12/c\mathcal{A}.$  As shown in Wolf {\it et al.}~\cite{Wolf}, considering the mutual information between subsystems leads to the conclusion that the area law of entropy emerges in the system with a short-range correlation, {\it i.e.}  $\xi$ is small, while the area law is violated as $\xi$ becomes infinite. In this way, the logarithmic scaling of black hole's entropy from the R\'enyi statistics as $r_h / L_\lambda \to \infty$, corresponding to $\xi/a \to \infty$, may imply the existence of long-range correlations between the black hole and its surounding in the large black hole limit.  It is interesting to note  that the nonextensivity length scale $L_\lambda$ works as a lattice spacing in spacetime, which should be larger than the Planck length $l_P$ as discussed in \ref{planck length}.  Therefore, the role of $L_\lambda$ may be a kind of the fine-graining or coarse-graining parameters for the entropy.   Remarkably, these tend to indicate that the role of correlations in black hole thermodynamics emerging in the R\'enyi thermodynamic description may be originated from some microscosopic aspects, which is interesting to be explored further.

Recently, there are many attempts to test classical general relativity and black hole thermodynamics, e.g., the gravitationnal wave experiments~\cite{Brustein, Carullo} and analogue gravity systems~\cite{Unruh, Svidzinsky, Mannarrelli}, also see~\cite{Barcelo} for a review. It may be worth in some ways to explore these phenomena using the R\'enyi statistics and investigate whether there is any deviation from the standard GB statistics.  This could indicate the nature of correlations in the system.

\section*{Acknowledgement}
We are grateful to Supakchai Ponglertsakul, Krai Cheamsawat and Tanapat Deesuwan for helpful discussions. This publication has been supported by the Petchra Pra Jom Klao Ph.D. Research Scholarship from King Mongkut's University of Technology Thonburi (KMUTT) and the basic research fund 2021 from Thailand Science Research and Innovation (TSRI).

\end{document}